\documentclass[aps,pra,twocolumn,floatfix,bibnotes]{revtex4-2}
\usepackage{graphicx}
\usepackage{amsmath}
\usepackage{amssymb}
\usepackage{bm}
\usepackage{color}
\usepackage{dcolumn}
\usepackage{scalerel}
\usepackage{multirow}
\usepackage[caption=false]{subfig}
\usepackage{hyperref}
\usepackage{float}

\begin{document}

\title{Classical-limit formula for matrix elements between bound states of distinct one-dimensional potentials}

\newcommand{\NIST}{
National Institute of Standards and Technology, Boulder, Colorado 80305, USA
}

\author{K. Beloy}
\email{kyle.beloy@nist.gov}
\affiliation{\NIST}

\date{\today}

\newcommand{\later}{\ensuremath{\spadesuit}}
\newcommand{\pprime}{\ensuremath{{\prime\prime}}}

\begin{abstract}
A classical-limit formula is derived for matrix elements between bound states of distinct one-dimensional potentials. We leave open the physical interpretation of the potentials, but generally envision them to be Born-Oppenheimer-type potentials with the position variable $x$ identifying with a slowly evolving degree of freedom of the system. For instance, $x$ could represent the internuclear separation in a diatomic molecule, with the potentials being potential energy curves for different electronic states. In this scenario, the matrix elements could be, e.g., conventional Franck-Condon factors. To test the derived formula, we assume functional forms for the potentials and operator that afford analytical solutions for the matrix elements. As the classical limit is approached, the computed matrix elements exhibit a clear tendency towards the classical-limit formula, providing strong validation for the formula. In future work, we anticipate using the formula to model inhomogeneous excitation spectra of atoms in one-dimensional optical lattices, with an eye towards improved optical lattice clock performance.
\end{abstract}

\maketitle

\section{Introduction}

The Born-Oppenheimer approximation is a ubiquitous tool in quantum mechanics. While it is most commonly associated with molecular physics, it has much broader applicability. For instance, our group has invoked it to describe motional (i.e., external-degree-of-freedom) states of atoms in one-dimensional optical lattices, with the axial and radial degrees of freedom playing the respective roles of the electronic and nuclear degrees of freedom in the molecular problem~\cite{BelMcGZha20}. This theory can be extended to improve modeling of inhomogeneous excitation spectra of lattice-trapped atoms~\cite{BlaThoCam09,GotPetCon25arXiv}. Such spectra contain information about the population distribution of the motional states, which is of critical importance to state-of-the-art optical lattice clocks~\cite{BroPhiBel17,UshTakKat18,NemJorYan19,BelMcGZha20}. Due to weak radial confinement, the density of motional states is high, and a challenge that arises is finding the relative transition strengths between motional states that are part of an effective continuum of states. This challenge has inspired the present work, and we anticipate that the results can ultimately be applied to the benefit of optical lattice clocks. However, here we present the problem in more generic terms, as the results could have broader applicability.

In this work, we derive a classical-limit formula for matrix elements between bound states of distinct one-dimensional potentials. Hereafter we refer to this classical-limit formula as the CLF. Generally, the potentials are envisioned to be Born-Oppenheimer-type potentials (e.g., potential energy curves attributed to different electronic states of a diatomic molecule). However, even this interpretation is not strictly necessary. For instance, one may be interested in a confining potential that changes abruptly at some point in time, with the ``before'' and ``after'' forms of the potential being taken as the distinct potentials in our analysis. In any case, the generic problem is outlined in the following section, while leaving open the physical interpretation of the potentials.

We test the CLF by assuming specific functional forms for the potentials and operator that afford analytical solutions for the matrix elements, allowing us to calculate the matrix elements with negligible numerical error. As the classical limit is approached, we observe a tendency towards the CLF, largely validating the formula.

\section{Problem set-up}

\begin{figure}[t]
\includegraphics[width=246pt]{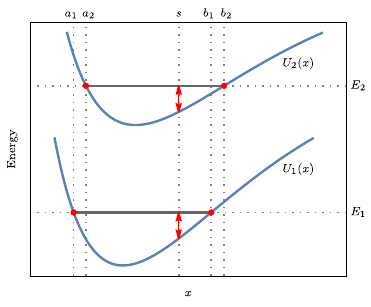}
\caption{Potentials $U_1(x)$ and $U_2(x)$ and bound-state energy levels $E_1$ and $E_2$ associated with the respective potentials. The points $a_1$, $b_1$, $a_2$, and $b_2$ are the CTPs [see Eq.~(\ref{Eq:CTP})], indicated by circles. The point $s$ is the COP [see Eq.~(\ref{Eq:COP})], indicated by arrows of equal length.}
\label{Fig:genericpotentials}
\end{figure}

We consider two one-dimensional potentials $U_i(x)$, with $i=1,2$. The potentials are assumed to vary slowly with $x$, with each potential supporting a high density of bound states. We will consider a pair of bound states, with one state associated with each potential. The bound states satisfy the Schr\"{o}dinger equation
\begin{gather*}
\left[-\frac{\hbar^2}{2m}\nabla^2+U_i(x)\right]\psi_i(x)=E_i\psi_i(x),
\end{gather*}
where $\hbar$ is the reduced Planck's constant and $m$ has dimensions of mass. Here $E_i$ and $\psi_i(x)$ are the energy and wave function for the bound state of interest associated with the potential $U_i(x)$. The wave functions are taken to be real and normalized. We assume $n_i\gg1$, where $n_i$ identifies with the number of nodes of the wave function $\psi_i(x)$ (i.e., the ground state has $n_i=0$ nodes, the first-excited state has $n_i=1$ node, etc.). Figure~\ref{Fig:genericpotentials} depicts the potentials and energy levels.

For convenience, we assume the potential difference $U_1(x)-U_2(x)$ varies monotonically with respect to $x$, with the two possible scenarios designated as ``case I'' and ``case II.'' Specifically,
\begin{gather*}
\begin{array}{ll}
\text{case I:}		&\quad U_1^\prime(x)>U_2^\prime(x)	\\
\text{case II:}		&\quad U_1^\prime(x)<U_2^\prime(x)
\end{array},
\end{gather*}
where primes are used throughout to denote derivatives with respect to $x$. While this assumption simplifies some of the theory, essential aspects of the theory can be applied more generally, as discussed in Appendix~\ref{Sec:extensions}. The potentials illustrated in Figure~\ref{Fig:genericpotentials} are case I. 

In this work, we concern ourselves with a ``matrix element'' between the pair of bound states. The matrix element is taken as
\begin{gather}
M=\int\psi_1(x)f(x)\psi_2(x)dx,
\label{Eq:mel}
\end{gather}
where the operator function $f(x)$ is some slowly varying function of $x$. The absence of explicit integration limits indicates integration over all physically acceptable values of $x$. To facilitate our analysis, we introduce the classical turning points (CTPs) $a_i$ and $b_i$ satisfying the condition
\begin{gather}
U_i\left(a_i\right)=U_i\left(b_i\right)=E_i,
\label{Eq:CTP}
\end{gather}
with $a_i<b_i$. In addition to the CTPs, we introduce a ``common oscillation point'' (COP), denoted by $s$. The COP satisfies the condition
\begin{gather}
E_1-U_1(s)=E_2-U_2(s),
\label{Eq:COP}
\end{gather}
with $a_1,a_2<s<b_1,b_2$. Given our assumption that the potential difference varies monotonically with $x$, a single COP exists provided the following criteria are met
\begin{gather}
\begin{array}{ll}
\text{case I:}		&\quad a_1<a_2<b_1<b_2	\\
\text{case II:}		&\quad a_2<a_1<b_2<b_1
\end{array}.
\label{Eq:CTPcriteria}
\end{gather}
These criteria can alternatively be expressed as
\begin{gather}
\begin{array}{ll}
\text{case I:}		&\quad U_1(a_2)<E_1<U_1(b_2)	\\
\text{case II:}		&\quad U_1(b_2)<E_1<U_1(a_2)
\end{array},
\label{Eq:E1limits}
\end{gather}
or
\begin{gather}
\begin{array}{ll}
\text{case I:}		&\quad U_2(b_1)<E_2<U_2(a_1)	\\
\text{case II:}		&\quad U_2(a_1)<E_2<U_2(b_1)
\end{array}.
\label{Eq:E2limits}
\end{gather}
No COP exists otherwise. Per our derivation below, the matrix element is assigned a value of zero unless a COP exists. Unless stated otherwise, we will therefore assume the criteria are met. The CTPs and the COP are depicted in Fig.~\ref{Fig:genericpotentials}.

To derive an expression for $M$, we leverage principles of the WKB approximation~\cite{Merzbacher}. The WKB approximation has its limitations. For instance, the WKB wave functions, given by Eq.~(\ref{Eq:WKBwavefunctions}) below, become invalid as $x$ approaches the CTPs. Ultimately, we regard our results as applying to the classical limit, where the bound-state energy spectrum of each potential approaches a continuum and limitations of the WKB approximation, as well as other approximations we employ, effectively drop away (e.g., the WKB wave functions become valid arbitrarily close to the CTPs, with negligible amplitude outside the classically allowed region). With this in mind, we proceed through our derivation without addressing conditions of validity associated with each step. Likewise, we will be content with mathematically imprecise language, such as characterizing functions as either ``slowly varying'' or ``rapidly oscillating'' with $x$. 

\section{Derivation of the CLF}

We begin by introducing the functions
\begin{gather}
k_i(x)=
\sqrt{\frac{2m}{\hbar^2}\left[E_i-U_i(x)\right]},
\label{Eq:k}
\end{gather}
valid within the classically allowed region $a_i\leq x\leq b_i$. Within the framework of the WKB approximation, $k_i(x)$ identifies with the local (angular) wavenumber for the wave function $\psi_i(x)$. In terms of $k_i(x)$, the defining conditions for the CTPs and COP, Eqs.~(\ref{Eq:CTP}) and (\ref{Eq:COP}), are
\begin{gather*}
k_i(a_i)=k_i(b_i)=0,
\\
k_1(s)=k_2(s).
\end{gather*}
Thus, the wave functions $\psi_1(x)$ and $\psi_2(x)$ have a common local wavenumber at the COP. We further introduce phase functions $\Phi_i(x)$ according to the derivative relation
\begin{gather*}
\Phi_i^\prime(x)=k_i(x),
\end{gather*}
together with the boundary conditions
\begin{gather*}
\Phi_i(a_i)=0,
\\
\Phi_i(b_i)=\pi\left(n_i+\frac{1}{2}\right).
\end{gather*}
These expressions for $\Phi_i(x)$ imply
\begin{gather}
\int_{a_i}^{b_i}k_i(x)dx=\pi\left(n_i+\frac{1}{2}\right),
\label{Eq:WKBenergy}
\end{gather}
which is the familiar WKB condition for bound-state energies. From equations~(\ref{Eq:k}) and (\ref{Eq:WKBenergy}), we infer the result (see Appendix~\ref{Sec:WKBdosderive} for additional details)
\begin{gather}
g_i=\frac{1}{2\pi}\frac{2m}{\hbar^2}\int_{a_i}^{b_i}
\frac{1}{k_i(x)}dx,
\label{Eq:dos}
\end{gather}
where $g_i$ is the density of states for the potential $U_i(x)$ at the energy $E_i$.

According to the WKB approximation, the wave functions in the classically allowed region are given by
\begin{gather}
\psi_i(x)\approx\frac{\mathcal{A}_i}{\sqrt{k_i(x)}}
\cos\left[\Phi_i(x)-\frac{\pi}{4}\right],
\label{Eq:wfA}
\end{gather}
where $\mathcal{A}_i$ is a (real) normalization constant. The cosine term oscillates rapidly in $x$, while the factor in front provides a slowly varying amplitude. We obtain an expression for the normalization constant by evaluating
\begin{gather*}
\int
\left[\psi_i(x)\right]^2dx
\approx
\mathcal{A}_i^2\int_{a_i}^{b_i}
\frac{1}{k_i(x)}
\cos^2\left[\Phi_i(x)-\frac{\pi}{4}\right]dx.
\end{gather*}
Replacing the rapidly oscillating cosine-squared factor with an average value of 1/2, this becomes
\begin{gather}
\int
\left[\psi_i(x)\right]^2dx
\approx
\pi\frac{\hbar^2}{2m}g_i\mathcal{A}_i^2,
\label{Eq:normset}
\end{gather}
where we have used Eq.~(\ref{Eq:dos}) for the density of states. We can now solve for $\mathcal{A}_i$ by equating the right-hand-side of Eq.~(\ref{Eq:normset}) to unity. Inserting the result into Eq.~(\ref{Eq:wfA}) leads to the following expression for the wave functions
\begin{gather}
\psi_i(x)\approx
\sqrt{\frac{1}{\pi}\frac{2m}{\hbar^2}\frac{1}{g_ik_i(x)}}
\cos\left[\Phi_i(x)-\frac{\pi}{4}\right].
\label{Eq:WKBwavefunctions}
\end{gather}

Using Eq.~(\ref{Eq:WKBwavefunctions}) for the wave functions, we can now write the matrix element as
\begin{gather}
M\approx
\frac{1}{\pi}\frac{2m}{\hbar^2}
\frac{1}{\sqrt{g_1g_2}}
\int
P(x)O(x)dx,
\label{Eq:Mfirstapprox}
\end{gather}
where the functions $P(x)$ and $O(x)$ are given by
\begin{gather*}
P(x)=\frac{f(x)}{\sqrt{k_1(x)k_2(x)}},
\\
O(x)=
\cos\left[\Phi_1(x)-\frac{\pi}{4}\right]
\cos\left[\Phi_2(x)-\frac{\pi}{4}\right].
\end{gather*}
The function $P(x)$ varies slowly with $x$. Using trigonometric identities, we express $O(x)$ as
\begin{gather}
O(x)=
\frac{1}{2}\sin\left[\Phi_1(x)+\Phi_2(x)\right]
+
\frac{1}{2}\cos\left[\Phi_1(x)-\Phi_2(x)\right].
\label{Eq:Orecast}
\end{gather}
The first term on the right-hand-side of Eq.~(\ref{Eq:Orecast}) oscillates rapidly in $x$ and averages to zero over short intervals. This is also the case for the second term, except in the immediate vicinity of the COP. A second-order expansion of the phase difference $\Phi_1(x)-\Phi_2(x)$ about the COP is
\begin{align*}
\Phi_1(x)-\Phi_2(x)
\approx
{}&
\left[\Phi_1(s)-\Phi_2(s)\right]
\\&
+\frac{1}{2}
\left[\Phi_1^\pprime(s)-\Phi_2^\pprime(s)\right](x-s)^2.
\end{align*}
There is no first-order term in this expansion because $\Phi_1^\prime(s)-\Phi_2^\prime(s)=k_1(s)-k_2(s)=0$. Using $\Phi_i^\pprime(x)=k_i^\prime(x)$ and Eq.~(\ref{Eq:k}) for $k_i(x)$, this expansion can be written as
\begin{align}
\Phi_1(x)-\Phi_2(x)
\approx
{}&
\left[\Phi_1(s)-\Phi_2(s)\right]
\nonumber\\&
-\frac{1}{4}\frac{2m}{\hbar^2}
\frac{1}{k(s)}\left[U_1^\prime(s)-U_2^\prime(s)\right](x-s)^2.
\label{Eq:Phidiffexpand}
\end{align}
Here we have dropped the index on $k(s)$, since the relation $k_1(s)=k_2(s)$ makes the index superfluous.

At this point, we are prompted to consider the general integral result
\begin{gather}
\int_{-\infty}^{+\infty}\cos\left(\beta\pm u^2\right)du=\sqrt{\pi}\cos\left(\beta\pm\frac{\pi}{4}\right).
\label{Eq:genint}
\end{gather}
For $|u|\gg1$, the integrand oscillates rapidly with respect to $u$ and averages to zero over small intervals. Consequently, the integral is predominantly accumulated in the neighborhood of $u\approx0$, though the infinite limits yield the simple analytical result here.

We now have the ingredients necessary to arrive at an expression for $M$. This is accomplished by taking the following steps: i) evaluate the slowly varying function $P(x)$ in Eq.~(\ref{Eq:Mfirstapprox}) at the COP and pull it out of the integral, ii) retain only the second term on the right-hand-side of Eq.~(\ref{Eq:Orecast}), iii) use the second-order expansion of Eq.~(\ref{Eq:Phidiffexpand}) for the phase difference $\Phi_1(x)-\Phi_2(x)$, and iv) use Eq.~(\ref{Eq:genint}) to evaluate the remaining integral of Eq.~(\ref{Eq:Mfirstapprox}), which is accumulated predominantly in the vicinity of the COP. Following these steps, we arrive at the result
\begin{align}
M\approx
{}&
\sqrt{\frac{1}{\pi}\frac{2m}{\hbar^2}}
\frac{f(s)}{\sqrt{g_1g_2k(s)\left|U_1^\prime(s)-U_2^\prime(s)\right|}}
\nonumber\\&\times
\cos\left[\Phi_1(s)-\Phi_2(s)\mp\frac{\pi}{4}\right],
\label{Eq:Mapproxcosine}
\end{align}
where the top (bottom) sign applies for case I (II).

We take additional steps aimed at the cosine factor in Eq.~(\ref{Eq:Mapproxcosine}). We first acknowledge that the overall sign of the wave functions is arbitrary, with the sign of $M$ likewise being arbitrary. We thus concern ourselves with $\left|M\right|$ in what follows. Moreover, one may be interested in $\left|M\right|$ raised to some power $r$, where we assume $r>0$. Finally, as the classical limit is approached---and the bound-state spectrum for each potential approaches a continuum---the quantity $\left|M\right|^r$ associated with a specific pair of bound states generally loses significance. Rather, we concern ourselves with $\left|M\right|^r$ averaged over states that have energies in the immediate vicinity of $E_1$ and $E_2$, with $E_1$ and $E_2$ now being regarded as continuous energy variables rather than identifying with specific bound states. This averaging procedure effectively randomizes the phase quantity $\left[\Phi_1(s)-\Phi_2(s)\mp\frac{\pi}{4}\right]$ mod $\pi$. To account for these considerations, we make the following substitution
\begin{align*}
\left|\cos\left[\Phi_1(s)-\Phi_2(s)\pm\frac{\pi}{4}\right]\right|^r
&\rightarrow
\frac{1}{2\pi}\int_{0}^{2\pi}\left|\cos u\right|^rdu
\\&
=\frac{\Gamma\left(\frac{1+r}{2}\right)}
{\sqrt{\pi}\Gamma\left(1+\frac{r}{2}\right)},
\end{align*}
where $\Gamma(z)$ is the conventional $\Gamma$-function. As specific examples, the expression on the second line evaluates to $2/\pi$ and 1/2 for $r$ equal to 1 and 2, respectively. Thus, Eq.~(\ref{Eq:Mapproxcosine}) becomes
\begin{align}
\left|M\right|^r
={}&
\frac{\Gamma\left(\frac{1+r}{2}\right)}
{\sqrt{\pi}\Gamma\left(1+\frac{r}{2}\right)}
\left|f(s)\right|^r
\nonumber\\&\times
\left[
\pi\frac{\hbar^2}{2m}
g_1g_2k(s)\left|U_1^\prime(s)-U_2^\prime(s)\right|
\right]^{-r/2},
\label{Eq:Mapprox2}
\end{align}
where the averaging procedure is now implicit. We refer to this result as the CLF.

With $E_1$ and $E_2$ now regarded as continuous variables, $\left|M\right|^r$ is therefore an implicit function of these variables. The densities of states $g_1$ and $g_2$ are implicit functions of their respective energies $E_1$ and $E_2$. From Eq.~(\ref{Eq:COP}), we infer that $s$ is an implicit function of the energy difference $E_1-E_2$. It follows that the quantities $f(s)$ and $\left|U_1^\prime(s)-U_2^\prime(s)\right|$ are also implicit functions of this energy difference. Finally, accounting for the explicit energy appearing on the right-hand-side of Eq.~(\ref{Eq:k}), the remaining factor $k(s)$ in Eq.~(\ref{Eq:Mapprox2}) is an implicit function of both $E_1$ and $E_2$. These dependencies are exemplified in Sec.~\ref{Sec:expressCLF} below, where specific functional forms of the potentials and operator are assumed.

\section{Testing the CLF with $f(x)=1$ and $r=2$}

Here we consider the specific case of $f(x)=1$ and $r=2$, as it allows a check of the CLF. For $f(x)=1$, the matrix element $M$ amounts to the overlap integral between wave functions. If we square this matrix element and sum over all states of either potential, we expect the result to be unity. Summing over all states of potential $U_2(x)$, for instance, we expect
\begin{gather}
\sum_{n_2}\left|M\right|^2=1,
\label{Eq:sumn2}
\end{gather}
where $n_2$ is used to index the states of potential $U_2(x)$ (see additional comments in Appendix~\ref{Sec:potcomments}). Regarding $\left|M\right|^2$ to be a function of the continuous variables $E_1$ and $E_2$, the summation over $n_2$ is replaced by an integration over $E_2$. The number of states within a differential energy interval $dE_2$ at $E_2$ is $g_2dE_2$, where we recall that the density of states $g_2$ is an implicit function of $E_2$. The integration limits for $E_2$ follow from expression~(\ref{Eq:E2limits}). Thus, the discrete summation $\sum_{n_2}\left|M\right|^2$ is replaced with the integral
\begin{gather}
I=\mp\int_{U_2(a_1)}^{U_2(b_1)}
\left|M\right|^2
g_2dE_2,
\label{Eq:Ifirst}
\end{gather}
where the top (bottom) sign applies for case I (II). The negative sign for case I accounts for the inverted ordering of the integration limits relative to expression~(\ref{Eq:E2limits}).

Using Eq.~(\ref{Eq:Mapprox2}) with $f(x)=1$ and $r=2$, the integral becomes
\begin{gather*}
I=\mp
\frac{1}{2\pi}\frac{2m}{\hbar^2}\frac{1}{g_1}
\int_{U_2(a_1)}^{U_2(b_1)}
\frac{1}{k(s)\left|U_1^\prime(s)-U_2^\prime(s)\right|}
dE_2.
\end{gather*}
The $\left|M\right|^2$ in Eq.~(\ref{Eq:Ifirst}) brings factors of $1/g_1$ and $1/g_2$. The factor $1/g_1$ is brought outside of the integral here because it is an implicit function of $E_1$ only. Meanwhile, the factor $1/g_2$ cancels with the explicit $g_2$ from Eq.~(\ref{Eq:Ifirst}). At this point, we may perform a change of variables, taking $s$ as the integration variable rather than $E_2$. From Eq.~(\ref{Eq:COP}) with $E_1$ treated as fixed, we find
\begin{gather*}
dE_2=\left[U_2^\prime(s)-U_1^\prime(s)\right]ds.
\end{gather*}
This can be rewritten as
\begin{gather*}
dE_2=\mp\left|U_1^\prime(s)-U_2^\prime(s)\right|ds,
\end{gather*}
where the top (bottom) sign applies for case I (II). With this change of variables from $E_2$ to $s$, the lower and upper limits transform as $U_2(a_1)\rightarrow a_1$ and $U_2(b_1)\rightarrow b_1$, respectively. Using $k(s)=k_1(s)$, the integral becomes
\begin{gather*}
I=
\frac{1}{2\pi}\frac{2m}{\hbar^2}\frac{1}{g_1}
\int_{a_1}^{b_1}
\frac{1}{k_1(s)}
ds,
\end{gather*}
which applies for both case I and case II. Using Eq.~(\ref{Eq:dos}) for the density of states, we arrive at the expected result $I=1$.

\section{Testing the CLF with analytical results}
\label{Sec:analytical}

Here we consider functional forms of $U_i(x)$ and $f(x)$ that afford analytical expressions for the matrix elements, as well as quantities appearing in the CLF. This allows a further check of the CLF. Our choice of potentials is partially motivated by a physical problem of interest. We reserve those details for the Conclusion, as they are not crucial for the goals of the present section.

The potentials are taken to have the form
\begin{gather}
U_i(x)=
\frac{A}{\mathrm{sinh}^2\left(\alpha x\right)}
-\frac{B_i}{\mathrm{cosh}^2\left(\alpha x\right)},
\label{Eq:SPT}
\end{gather}
where $\alpha$ has dimensions of inverse length and $A$ and $B_i$ have dimensions of energy. Here $\alpha$ and $A$ are taken to be common between the potentials, while $B_1$ and $B_2$ are distinct. We express the parameters $(\alpha,A,B_i)$ in terms of a second set of parameters $(\epsilon,p,\nu_i)$ as
\begin{gather*}
\alpha=\sqrt{2m\epsilon/\hbar^2},
\\
A=\epsilon\left(p^2-1/4\right),
\\
B_i=\epsilon\left[\left(2\nu_i+p+1\right)^2-1/4\right],
\end{gather*}
where $\epsilon$ has dimensions of energy and $p$ and $\nu_i$ are dimensionless. For our purposes, we take $p$ and $\nu_i$ to be integers satisfying $p,\nu_i\gg1$. We will regard the parameters $(\epsilon,p,\nu_i)$ as more fundamental, with $(\alpha,A,B_i)$ being used for conciseness as appropriate. The particle is constrained to the region $x>0$, with $U_i(x)\rightarrow\infty$ as $x\rightarrow0$ and $U_i(x)\rightarrow0$ as $x\rightarrow\infty$ (see additional comments in Appendix~\ref{Sec:potcomments}). We restrict our attention to the bound states, which have energies $E_i<0$.

With the potentials given by Eq.~(\ref{Eq:SPT}), the  one-dimensional time-independent Schr\"odinger equation reduces to the second P\"oschl-Teller differential equation. Analytical solutions exist for the bound-state energies and wave functions~\cite{YouLuSun13,Younote}. The energies are given by
\begin{gather}
E_i=-4\epsilon\left(\nu_i-n_i\right)^2,
\label{Eq:SPTenergies}
\end{gather}
where the allowed values of $n_i$ are
\begin{gather*}
n_i=0,1,2,\dots,\nu_i-1.
\end{gather*}
The parameter $\nu_i$ therefore identifies with the number of bound states supported by the potential $U_i(x)$.

The operator function $f(x)$ is taken to have the form
\begin{gather*}
f(x)=\frac{1}{\cosh^{2t}(\alpha x)}.
\end{gather*}
Below we consider the cases $t=0$ and $t=1$. The former corresponds to $f(x)=1$, with the matrix elements amounting to overlap integrals between the wave functions. The latter provides a test of $f(x)\neq1$.

We associate the classical limit with $\sigma\rightarrow\infty$ and fixed ratios $p{\,:\,}\nu_1{\,:\,}\nu_2$, where $\sigma$ represents any of the three parameters ($p,\nu_1,\nu_2$). We further stipulate that the energy quantities $\epsilon \sigma^2$ remain constant as $\sigma\rightarrow\infty$. This implies that $A$, $B_1$, and $B_2$ approach constant values. It also implies that $\epsilon$ approaches zero, with the bound-state energy spectrum for each potential becoming infinitely dense. It is worth noting that, while we find it conceptually convenient to assume that $\epsilon \sigma^2$ is constant as $\sigma\rightarrow\infty$, so as to connect with the conventional notion of the classical limit, expressions to follow have no reliance on this assumption. Similarly, given the relation $\epsilon=\hbar^2\alpha^2/2m$, one may find it conceptually convenient to further attribute the classical limit to, e.g., $\hbar\rightarrow0$, $\alpha\rightarrow0$, or $m\rightarrow\infty$.

\subsection{Analytical expression for the CLF}
\label{Sec:expressCLF}

Given the above forms of $U_i(x)$ and $f(x)$, here we derive an analytical expression for the CLF, presented as a function of the energy variables $E_1$ and $E_2$. We start by addressing the density of states $g_i$. An expression for $g_i$ may be obtained from Eqs.~(\ref{Eq:k}) and (\ref{Eq:dos}), as demonstrated in Appendix~\ref{Sec:dosWKBSPT}. A more efficient approach exploits the known energy spectrum, Eq.~(\ref{Eq:SPTenergies}), together with the relation $g_i=\left(dE_i/dn_i\right)^{-1}$. Expressing the result in terms of the energy $E_i$, we have
\begin{gather*}
g_i=\frac{1}{4\sqrt{-\epsilon E_i}}.
\end{gather*}

Next, we turn to the remaining factors appearing in the CLF. Equation~(\ref{Eq:COP}) for the COP implies
\begin{gather*}
\cosh(\alpha s)=\sqrt{\frac{B_1-B_2}{E_2-E_1}}.
\end{gather*}
Allowed values of $E_1$ and $E_2$ are discussed below; for now we merely acknowledge the criterion $0<\left(E_2-E_1\right)/\left(B_1-B_2\right)<1$ for a COP to exist. The above result is useful for obtaining expressions for $f(s)$, $U_1^\prime(s)-U_2^\prime(s)$, and $k(s)$. For instance, $f(s)$ is readily found to be
\begin{gather*}
f(s)=
\left(
\frac{E_2-E_1}{B_1-B_2}
\right)^{t}.
\end{gather*}
With a little more effort, we also find
\begin{gather*}
U_1^\prime(s)-U_2^\prime(s)
=2\alpha\left(E_2-E_1\right)\left(1-\frac{E_2-E_1}{B_1-B_2}\right)^{1/2},
\\
k(s)=\sqrt{\frac{2mW}{\hbar^2}}\left(1-\frac{E_2-E_1}{B_1-B_2}\right)^{-1/2},
\end{gather*}
where we have introduced
\begin{gather}
W=-A\frac{E_2-E_1}{B_1-B_2}
+\left(1-\frac{E_2-E_1}{B_1-B_2}\right)
\frac{B_1E_2-B_2E_1}{B_1-B_2}.
\label{Eq:W}
\end{gather}
Inserting each of these results into the CLF, Eq.~(\ref{Eq:Mapprox2}), we have
\begin{align}
\left|M\right|^r=
{}&
\frac{\Gamma\left(\frac{1+r}{2}\right)}
{\sqrt{\pi}\Gamma\left(1+\frac{r}{2}\right)}
\left(\frac{8}{\pi}\right)^{r/2}
\left(\frac{E_2-E_1}{B_1-B_2}\right)^{rt}
\nonumber\\
&\times
\left[\frac{E_1E_2}{\left(E_2-E_1\right)^2}\frac{\epsilon}{W}\right]^{r/4}.
\label{Eq:MSPT}
\end{align}
Thus, we have our desired expression for $\left|M\right|^r$ as a function of the energies $E_1$ and $E_2$.

To close this section, we address the allowed values of $E_1$ and $E_2$. Independently, each $E_i$ is limited to the range $-D_i<E_i<0$, corresponding to bound states, where $D_i=\left(\sqrt{B_i}-\sqrt{A}\right)^2$ is the potential depth of $U_i(x)$. Within these ranges, combinations of $E_1$ and $E_2$ are restricted to those for which a COP exists. This occurs when $W>0$. Let us consider the allowed values of $E_2$ for a given value of $E_1$. To find the limiting values of $E_2$, we consider the two roots of Eq.~(\ref{Eq:W}) when $W$ is set to zero. The roots are real and identify with $U_2(a_1)$ and $U_2(b_1)$, with solutions given by
\begin{widetext}
\begin{gather*}
\left\{\begin{array}{c}
U_2\left(a_1\right)\\
U_2\left(b_1\right)
\end{array}\right\}
=\frac{1}{2B_1}
\left[
\left(B_1-A\right)\left(B_1-B_2\right)
+E_1\left(B_1+B_2\right)
\pm
\left(B_1-B_2\right)
\sqrt{E_1^2+2\left(B_1+A\right)E_1+\left(B_1-A\right)^2
}
\right].
\end{gather*}
\end{widetext}
The top and bottom sign on the right-hand side correspond to $U_2\left(a_1\right)$ and $U_2\left(b_1\right)$, respectively. An inequality $B_1>B_2$ implies $U_2(b_1)<U_2(a_1)$, whereas an inequality $B_2>B_1$ implies $U_2(a_1)<U_2(b_1)$. The former is case I, whereas the latter is case II. In either case, $W$ may be expressed as
\begin{gather*}
W=-\frac{B_1}{\left(B_1-B_2\right)^2}
\left[E_2-U_2\left(b_1\right)\right]\left[E_2-U_2\left(a_1\right)\right].
\end{gather*}
The values of $E_2$ that render positive $W$ are consistent with expression~(\ref{Eq:E2limits}).

\subsection{Analytical expression for the matrix elements}
\label{Sec:expressmels}

For our purposes, we assume that $p$ and $\nu_i$ are integers. An analytical expression for the matrix elements can be obtained even when $p$ and $\nu_i$ take on non-integer values. However, the expression involves functions that generally evaluate to irrational values (e.g., Gamma functions), as well as summations over many terms with a high degree of cancellation between positive and negative contributions. This opens the door to numerical error in the evaluation of the matrix elements. By stipulating that $p$ and $\nu_i$ are integers, the expressions can be cast in terms of functions that evaluate to rational numbers (e.g., binomial coefficients). While the expression still involves summations over many terms, with a high degree of cancellation between positive and negative contributions, the terms involved are rational numbers. This allows the summations to be evaluated exactly using commercial mathematical software~\cite{Mathematica}, effectively eliminating numerical error in the evaluation of the matrix elements.

The analytical solutions for the wave functions are
\begin{align}
\psi_i(x)={}&
\sqrt{2\alpha}N_i
\left[1-w(x)\right]^{\nu_i-n_i}
\nonumber\\&\times
\sum_{\mu=0}^{n_i}C_i(\mu)
\left[w(x)\right]^{p/2+1/4+\mu}
\label{Eq:wf1}
\end{align}
where $w(x)=\tanh^2\left(\alpha x\right)$ and where $N_i$ and $C_i(\mu)$ are given by
\begin{gather*}
N_i=\sqrt{
2\left(\nu_i-n_i\right)
\left(\begin{array}{c}
n_i+p\\
n_i
\end{array}\right)
\left(\begin{array}{c}
2\nu_i+p-n_i\\
2\nu_i-n_i
\end{array}\right)
},
\\
C_i(\mu)=
(-1)^{\mu}
\frac{
\left(\begin{array}{c}n_i\\\mu\end{array}\right)
\left(\begin{array}{c}2\nu_i-n_i+p+\mu\\\mu\end{array}\right)
}{
\left(\begin{array}{c}p+\mu\\\mu\end{array}\right)
}.
\end{gather*}
Noting the relation $f(x)=\left[1-w(x)\right]^{t}$ and the derivative relation $w^\prime(x)=2\alpha\left[1-w(x)\right]\left[w(x)\right]^{1/2}$, the matrix element can therefore be written
\begin{align*}
M={}&
N_1N_2
\sum_{\mu_1=0}^{n_1}C_1(\mu_1)
\sum_{\mu_2=0}^{n_2}C_2(\mu_2)
\\&\times
\int_0^1
\left(1-w\right)^{\nu_1-n_1+\nu_2-n_2+t-1}w^{p+\mu_1+\mu_2}
dw,
\end{align*}
where we used the derivative relation for $w(x)$ to change the integration variable from $x$ to $w$. Explicit limits are provided for $w$, which correspond to integration over all physically acceptable values of $x$ (i.e., $x>0$). Evaluating the integral, we get
\begin{gather}
M=
N_1N_2
\sum_{\mu_1=0}^{n_1}
\sum_{\mu_2=0}^{n_2}
\frac{C_1(\mu_1)C_2(\mu_2)}{Z\left(\mu_1,\mu_2\right)},
\label{Eq:Manalytical}
\end{gather}
where
\begin{align*}
Z\left(\mu_1,\mu_2\right)
={}&
\left(\nu_1-n_1+\nu_2-n_2+t\right)
\\
&\hspace{-22.5pt}\times
\left(\begin{array}{c}
\nu_1-n_1+\nu_2-n_2+t+p+\mu_1+\mu_2 \\
\nu_1-n_1+\nu_2-n_2+t
\end{array}\right).
\end{align*}

\subsection{Comparison}

\begin{figure*}[tb]
\subfloat{\includegraphics[width=163pt]{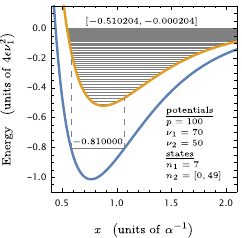}}%
\hspace{10.5pt}%
\subfloat{\includegraphics[width=163pt]{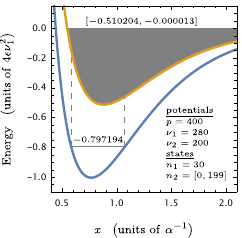}}%
\hspace{10.5pt}%
\subfloat{\includegraphics[width=163pt]{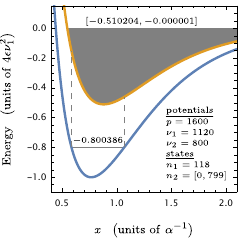}}%
\caption{Potentials and states used in the matrix element computations. The matrix elements are computed using analytical expressions given in Sec.~\ref{Sec:expressmels}. Results for the matrix elements are displayed in Figs.~\ref{Fig:melsA} and \ref{Fig:melsB} below. The parameters $p$, $\nu_1$, and $\nu_2$ are specified within the respective panel. The parameter ratios are the same across all three panels ($p{\,:\,}\nu_1{\,:\,}\nu_2=10{\,:\,}7{\,:\,}5$). From left to right, the panels represent a progression towards the classical limit. In each panel, curves depict the potentials $U_1(x)$ (blue, deeper) and $U_2(x)$ (yellow, shallower). For the states used in the matrix element computations, the energies are displayed as horizontal gray lines spanning between the respective CTPs. For potential $U_1(x)$, a single state is considered. It is taken to be the state with energy closest to $-0.8\times4\epsilon\nu_1^2$. For potential $U_2(x)$, all bound states are considered. The lines blend together where the density of states is high. The values for $n_1$ and $E_1$ and the range of values for $n_2$ and $E_2$ are provided within the respective panel. The chosen energy unit of $4\epsilon\nu_1^2$ corresponds to the approximate depth of $U_1(x)$. The vertical dashed lines align with the CTPs $a_1$ and $b_1$. These lines extend vertically from the energy value $U_1(a_1)=U_1(b_1)=E_1$ to the energy values $U_2(a_1)$ and $U_2(b_1)$, respectively.}
\label{Fig:potentialsandstates}
\end{figure*}

\begin{figure*}[tb]
\subfloat{\includegraphics[width=160pt]{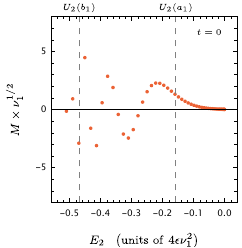}}%
\hspace{15pt}%
\subfloat{\includegraphics[width=160pt]{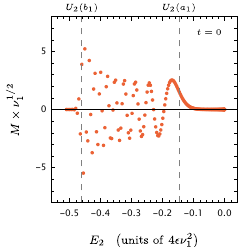}}%
\hspace{15pt}%
\subfloat{\includegraphics[width=160pt]{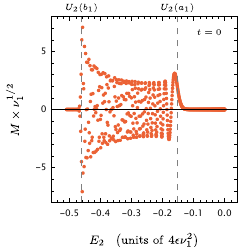}}%
\\
\subfloat{\includegraphics[width=160pt]{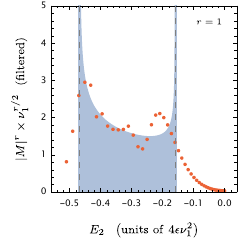}}%
\hspace{15pt}%
\subfloat{\includegraphics[width=160pt]{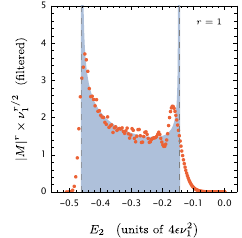}}%
\hspace{15pt}%
\subfloat{\includegraphics[width=160pt]{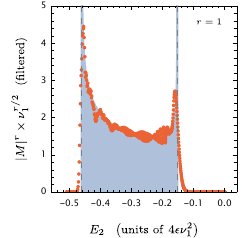}}%
\\
\subfloat{\includegraphics[width=160pt]{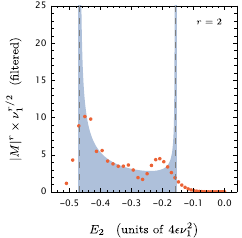}}%
\hspace{15pt}%
\subfloat{\includegraphics[width=160pt]{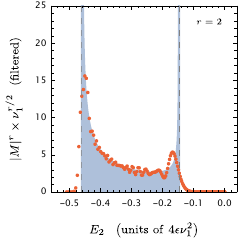}}%
\hspace{15pt}%
\subfloat{\includegraphics[width=160pt]{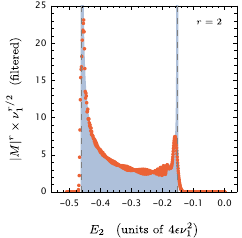}}%
\caption{Computed matrix elements with comparison to the CLF. These results are for $t=0$, corresponding to the function $f(x)=1$ (i.e., the matrix elements amount to the overlap integrals between wave functions). The states involved are specified in Fig.~\ref{Fig:potentialsandstates}, with each panel of Fig.~\ref{Fig:potentialsandstates} corresponding to a column of panels here. As in Fig.~\ref{Fig:potentialsandstates}, left to right represents a progression towards the classical limit. Top row: Results for the matrix elements $M$. Including a factor $\nu_1^{1/2}$ allows the results to be displayed on a common vertical scale across the row. Vertical dashed lines identify with $U_2(b_1)$ and $U_2(a_1)$, as labeled. Middle ($r=1$) and bottom ($r=2$) rows: Results for $\left|M\right|^r$ with a mean filter applied to smooth out rapid variations in $\left|M\right|^r$ versus $E_2$~\cite{Mathematica}. Including a factor $\nu_1^{r/2}$ allows the results to be displayed on a common vertical scale across each row. The vertical dashed lines identify with $U_2(b_1)$ and $U_2(a_1)$, as in the top row. The CLF, Eq.~(\ref{Eq:Mapprox2}), is presented as the boundary between shaded (below) and unshaded (above) regions. As the classical limit is approached, tendency towards the CLF is evident.
}
\label{Fig:melsA}
\end{figure*}

\begin{figure*}[tb]
\subfloat{\includegraphics[width=160pt]{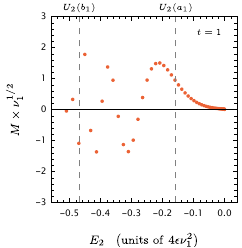}}%
\hspace{15pt}%
\subfloat{\includegraphics[width=160pt]{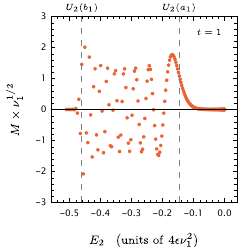}}%
\hspace{15pt}%
\subfloat{\includegraphics[width=160pt]{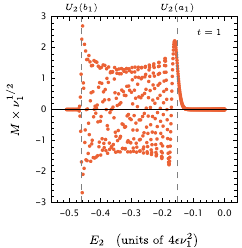}}%
\\
\subfloat{\includegraphics[width=160pt]{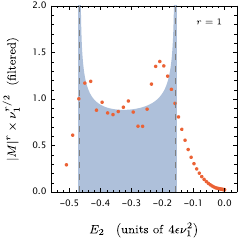}}%
\hspace{15pt}%
\subfloat{\includegraphics[width=160pt]{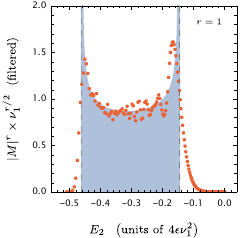}}%
\hspace{15pt}%
\subfloat{\includegraphics[width=160pt]{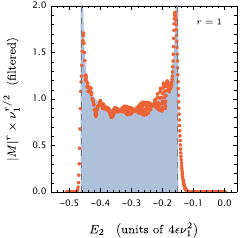}}%
\\
\subfloat{\includegraphics[width=160pt]{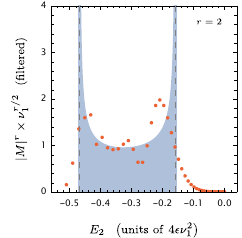}}%
\hspace{15pt}%
\subfloat{\includegraphics[width=160pt]{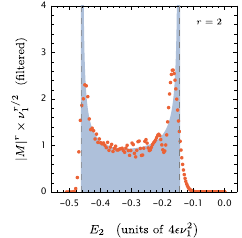}}%
\hspace{15pt}%
\subfloat{\includegraphics[width=160pt]{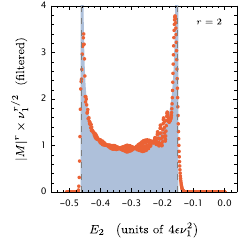}}%
\caption{Same as Figure~\ref{Fig:melsA}, except with $t=1$ rather than $t=0$.
}
\label{Fig:melsB}
\end{figure*}

Here we compare the results of Sec.~\ref{Sec:expressCLF} and \ref{Sec:expressmels}. Figure~\ref{Fig:potentialsandstates} illustrates the potentials and the states used for comparison purposes. The three panels of Figure~\ref{Fig:potentialsandstates}, from left to right, represent a progression towards the classical limit. Specifications for the potentials and states are provided within each panel. In each case, we consider a single bound state for potential $U_1(x)$ and all bound states for potential $U_2(x)$. Figure~\ref{Fig:melsA} presents computed matrix elements for the case $t=0$, corresponding to $f(x)=1$. Figure~\ref{Fig:melsA} is a $3\times3$ matrix of panels. The top row presents the matrix elements $M$ versus energy $E_2$. The second and third row present $\left|M\right|^r$ versus energy $E_2$ with $r=1$ and $r=2$, respectively, and with a mean filter applied to the data. The mean filter averages the values of $\left|M\right|^r$ among states having similar values of $E_2$. The filter bandwidth (in $E_2$ space) is taken to be a compromise between ``smoothing out'' fast oscillations in the data, while also allowing sharp ``edge'' features to be fairly well represented~\cite{Mathematica}. Also plotted in the second and third rows is the analytical result for the CLF. It is depicted as a boundary between shaded (below) and unshaded (above) regions within the plots. As the classical limit is approached (i.e., progressing from left to right across the panels), the matrix element data exhibits a clear trend towards the CLF. Figure~\ref{Fig:melsB} presents the same results as Figure~\ref{Fig:melsA}, except that $t=1$ is used rather than $t=0$. Here we also observe a clear trend towards the CLF as the classical limit is approached. Thus, these results provide strong validation of the CLF.

\section{Conclusion}

We have derived a classical-limit formula, Eq.~(\ref{Eq:Mapprox2}), for matrix elements between bound states of distinct one-dimensional potentials. To verify the formula, we assumed functional forms for the potentials and operator that afford analytical solutions for the matrix elements and quantities within the classical-limit formula. As the classical limit is approached, results for the matrix elements exhibit a clear tendency towards the classical-limit formula, largely validating the formula.

We anticipate using the classical-limit formula to improve modeling of inhomogeneous excitation spectra of atoms trapped in near-magic-wavelength optical lattices~\cite{BlaThoCam09,GotPetCon25arXiv}. Improved modeling of the inhomogeneous excitation spectra could lead to improved characterization of lattice light shifts in optical lattice clocks and, by consequence, improved optical lattice clock performance. It is worth noting that the potentials given by Eq.~(\ref{Eq:SPT}) are functionally similar to potentials that arise in this specific problem of interest, making the test comparisons here especially relevant. Namely, when the Born-Oppenheimer prescription of Ref.~\cite{BelMcGZha20} is invoked to describe atomic motion within the optical lattice, one obtains radial potential energy curves $U_{n_z}(\rho)$, where $\rho$ is the radial coordinate and the quantum number $n_z$ labels the state of axial motion. The next step involves solving a radial Schr\"odinger equation, which has the form of a one-dimensional Schr\"odinger equation with $\rho$ taking the place of $x$. For this step, the radial potentials $U_{n_z}(\rho)$ must be supplemented with a centrifugal term $(\hbar^2/2m)(l^2-1/4)\rho^{-2}$, where the quantum number $l$ is the angular momentum along the lattice axis in units of $\hbar$. Regarding the functional form of Eq.~(\ref{Eq:SPT}), we note that the three ``free'' parameters can be chosen to exactly match the lowest-order terms in $\rho$ of the true potential, including the centrifugal term ($\propto\rho^{-2}$), the harmonic term ($\propto\rho^2$), and the first-anharmonic term ($\propto\rho^4$), while also capturing additional anharmonic contributions (further note that a potential offset has no influence on the radial wave functions). Consequently, the functional form of Eq.~(\ref{Eq:SPT}) can provide a good representation of the radial potentials inclusive of the centrifugal term.

Beyond the anticipated application described above, the classical-limit formula derived here could find broader use, such as estimating Franck-Condon factors between certain electronic states of diatomic molecules.

\begin{acknowledgments}
The author thanks A.\ D.\ Ludlow and H.\ Ranganath for their careful reading of the manuscript. This work was supported by the National Institute of Standards and Technology/Physical Measurement Laboratory, an agency of the U.S.\ government, and is not subject to U.S.\ copyright. Data will be made available upon reasonable request.
\end{acknowledgments}

\appendix

\section{Relaxing the assumption that the potential difference varies monotonically with $x$}
\label{Sec:extensions}

For the purposes of the main text, we assume the potential difference $U_1(x)-U_2(x)$ varies monotonically with respect to $x$. As a consequence, there is at most a single COP for any combination of energies $E_1$ and $E_2$, expressions (\ref{Eq:CTPcriteria}), (\ref{Eq:E1limits}), and (\ref{Eq:E2limits}) apply for the existence of the COP, and $U_1^\prime(s)\neq U_2^\prime(s)$. For more general potentials, multiple COPs are possible. In this case, the CLF simply needs to be extended to include a summation over all COPs. For more general potentials, it is also possible that $U_1^\prime(s)=U_2^\prime(s)$ at the COP(s). This leads to a divergence in the CLF, which stems from the relation $\left[\Phi_1^\pprime(s)-\Phi_2^\pprime(s)\right]\propto\left[U_1^\prime(s)-U_2^\prime(s)\right]$. In this case, a second-order Taylor expansion of the phase difference $\left[\Phi_1(x)-\Phi_2(x)\right]$ about the COP is inadequate, with expansion to the leading nonvanishing order being necessary before proceeding with the derivation.

\section{Additional comments regarding the potentials}
\label{Sec:potcomments}

Nominally, we assume that the potentials only support bound states. However, this is not a critical assumption. As the classical limit is approached, the bound-state wave functions have negligible amplitude outside of the classically allowed region. In effect, the functional behavior of the potentials outside of the CTPs is of no consequence to the matrix elements. Consider the potentials and states used for comparison purposes in Sec.~\ref{Sec:analytical}. The potentials are open, such that the complete set of states associated with each potential includes both bound and unbound (continuum) states. The summation in Eq.~(\ref{Eq:sumn2}) should, in principle, be extended to include integration over the unbound states of potential $U_2(x)$. However, in the classical limit, these states do not contribute. We can use the matrix elements computed in Sec.~\ref{Sec:analytical} to illustrate this. For the case $f(x)=1$, the matrix elements are displayed in the top row of Fig.~\ref{Fig:melsA}, which includes all bound states of potential $U_2(x)$. We can square these matrix elements and add them up. The tallied result is shy of one, with the difference from one being the omitted continuum contribution. For the first, second, and third panels, the continuum contributions are found to be $1.5\times10^{-5}$, $6.8\times10^{-13}$, and $8.8\times10^{-44}$, respectively, exhibiting a steep reduction in the continuum contribution as the classical limit is approached. These values are computed exactly~\cite{Mathematica}, but rounded to two digits here. In reduced form, the exact rational numbers have denominators with 73, 298, and 1210 digits, respectively.

As mentioned above, in the classical limit, the behavior of the potentials outside of the CTPs is of no consequence. As such, the fact that the potentials used in Sec.~\ref{Sec:analytical} diverge to infinity as $x\rightarrow0$ is also of no consequence. More generally, the singularity can be treated with additional care within the WKB framework, applicable even to the case $p=0$~\cite{BerOzo73}.

\section{WKB expression for the density of states}
\label{Sec:WKBdosderive}

Here we derive the WKB expression for the density of states, Eq.~(\ref{Eq:dos}). For the purpose of this Appendix, we drop the potential index $i$ and include the energy $E$ as an explicit variable. Specifically, here we write the density of states as $g(E)$, the CTPs as $a(E)$ and $b(E)$, and the local wavenumber as $k\left(E,x\right)$.

We introduce the WKB phase $\phi(E)$ as
\begin{gather*}
\phi(E)=
\int_{a(E)}^{b(E)}k\left(E,x\right)dx.
\end{gather*}
The WKB condition for bound-state energies is
\begin{gather*}
\phi(E)=\pi\left(n+\frac{1}{2}\right),
\end{gather*}
where $n=0,1,2,\dots$ From this condition, we infer the density of states to be
\begin{gather*}
g(E)=\frac{1}{\pi}\frac{d}{dE}\phi(E).
\end{gather*}
The derivative of $\phi(E)$ is given by
\begin{align*}
\frac{d}{dE}\phi(E)=
&
\int_{a(E)}^{b(E)}
\left[
\frac{\partial}{\partial E}k\left(E,x\right)
\right]
dx
\\&
+\left[k\left(E,x\right)\right]_{x=b(E)}\left[\frac{d}{dE}b(E)\right]
\\&
-\left[k\left(E,x\right)\right]_{x=a(E)}\left[\frac{d}{dE}a(E)\right].
\end{align*}
Following from the definition of the CTPs, the last two terms on the right-hand-side vanish. Meanwhile, the form of $k\left(E,x\right)$, Eq.~(\ref{Eq:k}), implies
\begin{gather*}
\frac{\partial}{\partial E}k\left(E,x\right)
=\frac{1}{2}\frac{2m}{\hbar^2}
\frac{1}{k\left(E,x\right)}.
\end{gather*}
We therefore arrive at the result
\begin{gather*}
g(E)=\frac{1}{2\pi}\frac{2m}{\hbar^2}\int_{a(E)}^{b(E)}\frac{1}{k(E,x)}dx.
\end{gather*}
This is Eq.~(\ref{Eq:dos}), except with the energy variable explicit and the potential index dropped.

\section{Density of states for the potentials of Eq.~(\ref{Eq:SPT})}
\label{Sec:dosWKBSPT}

For the potentials given by Eq.~(\ref{Eq:SPT}), the density of states is most efficiently derived from the known energy spectrum. However, it is worth demonstrating that the WKB expression for the density of states, Eq.~(\ref{Eq:dos}), yields the same result. For the purpose of this Appendix, we drop the potential index $i$. The WKB expression for the density of states is
\begin{gather*}
g=\frac{1}{2\pi}\frac{2m}{\hbar^2}\int_a^b\frac{1}{k(x)}dx.
\end{gather*}
In present case, $k(x)$ is given by
\begin{gather*}
k(x)=\sqrt{\frac{2m}{\hbar^2}}
\sqrt{E-\left[\frac{A}{\sinh^2\left(\alpha x\right)}-\frac{B}{\cosh^2\left(\alpha x\right)}\right]}.
\end{gather*}
Using substitution with $u=\cosh^2\left(\alpha x\right)$, we find
\begin{gather*}
g=\frac{1}{4\pi\sqrt{\epsilon}}
\int_c^d
\frac{1}{\sqrt{
Eu(u-1)-Au+B(u-1)
}}
du,
\end{gather*}
where we have used $\epsilon=\hbar^2\alpha^2/2m$ and have introduced $c$ and $d$ as the integration limits for $u$. We recall that $E<0$ for the bound states. We further note that the quantity under the square root sign in the integral is a quadratic function of $u$. This quantity necessarily vanishes at the limits $c$ and $d$ and is positive for intermediate values of $u$. It follows that we may write
\begin{gather*}
g=\frac{1}{4\pi\sqrt{\epsilon}}
\int_c^d
\frac{1}{\sqrt{-E(u-c)(d-u)}}
du,
\end{gather*}
without making explicit reference to the potential parameters $A$ and $B$ within the integrand. The factor $1/\sqrt{-E}$ can be pulled out of the integral. For the remaining integral, we perform another substitution with $u=(d-c)\sin^2\theta+c$ to find
\begin{gather*}
\int_c^d
\frac{1}{\sqrt{(u-c)(d-u)}}
du
=
2\int_0^{\pi/2}
d\theta=\pi.
\end{gather*}
With this, we arrive at the final result
\begin{gather*}
g=\frac{1}{4\sqrt{-\epsilon E}},
\end{gather*}
which agrees with the result obtained from the known energy spectrum.


\begin{thebibliography}{11}
\bibitem [{You()}]{Younote}%
  \BibitemOpen
  \href@noop {} {}\bibinfo {note} {We note a typographical error in Eq.~(14) of
  Ref.~\cite{YouLuSun13}. The factor
  $\left(\lambda^\prime-\kappa^\prime-2n_r^2\right)$ in this equation should
  instead be $\left(\lambda^\prime-\kappa^\prime-2n_r\right)^2$.}\BibitemShut
  {Stop}%
\bibitem [{Mat()}]{Mathematica}%
  \BibitemOpen
  \href@noop {} {}\bibinfo {note} {We use Mathematica, version 13 (Wolfram
  Research{,} Inc.) to evaluate matrix elements $M$ given by
  Eq.~(\ref{Eq:Manalytical}) and for subsequent data processing in
  Figures~\ref{Fig:melsA} and \ref{Fig:melsB}. The mean filtering was performed
  with intrinsic Mathematica functions. Namely, the functions
  ``\mbox{TimeSeries}'' and ``\mbox{MeanFilter}'' were sequentially applied to
  lists containing data for $\left|M\right|^r$ versus $E_2$. The
  \mbox{MeanFilter} function takes an argument for the filter bandwidth, which
  we choose to be $\epsilon\nu_1^{3/2}$. Specification of commercial products
  is for informational purposes only. This does not imply recommendation or
  endorsement of any commercial product by NIST, nor does it imply that the
  products identified are necessarily the best available for these
  purposes.}\BibitemShut {Stop}%
\end{thebibliography}

%

\end{document}